\documentclass[fleqn,usenatbib,letters]{mnras}
\usepackage{newtxtext,newtxmath}

\usepackage[T1]{fontenc}

\DeclareRobustCommand{\VAN}[3]{#2}
\let\VANthebibliography\thebibliography
\def\thebibliography{\DeclareRobustCommand{\VAN}[3]{##3}\VANthebibliography}

\usepackage{graphicx}	
\usepackage{amsmath}	

\title[Relativistic blob in M87 jet]{Relativistic blob signatures in the M87 jet at sub-parsec scales}

\author[D. Mora et al.]{
Donaldo Mora,$^{1}$\thanks{E-mail: doemora@astro.unam.mx (DM)}
Alejandro Cruz-Osorio,$^{1}$\thanks{E-mail: aosorio@astro.unam.mx (ACO)}
Erika Benítez,$^{1}$\thanks{E-mail: erika@astro.unam.mx (EB)}
Yosuke Mizuno$^{2,3}$.
\\
$^{1}$Universidad Nacional Autónoma de México. Instituto de Astronomía. A.P. 70-264, 04510. Ciudad de México, México.\\
$^{2}$Tsung-Dao Lee Institute, Shanghai Jiao Tong University, Shanghai, 201210, People's Republic of China\\
$^{3}$School of Physics and Astronomy, Shanghai Jiao Tong University, Shanghai, 200240, People’s Republic of China
}

\date{Accepted XXX. Received YYY; in original form ZZZ}

\pubyear{\the\year{}}

\begin{document}
\label{firstpage}
\pagerange{\pageref{firstpage}--\pageref{lastpage}}
\maketitle

\begin{abstract}
In this work, we present results from general-relativistic radiative-transfer calculations of black hole accretion and jet launching to reproduce the observed properties of the M\,87 jet at sub-parsec scales. For the first time, in the simulations we have included a blob component at the launching region of the jet that allowed us to reproduce the quasi-simultaneous low-energy spectral energy distribution  ($10^{10} \text{ Hz} \leq \nu \leq 10^{16} \text{ Hz}$). Moreover, we obtain a better fit of the synchrotron emission in the self-absorption region ($\nu \leq 10^{11} \text{ Hz}$), showing that the inclusion of a blob component in our radiative transfer calculations results in a flatter spectrum ($\alpha_{22-86\ \text{GHz}}=0.08$). Additionally, our simulations improved the accuracy of the observed morphology of M\,87's jet, matching the data up to a distance of 0.6 mas from the core at 86 GHz. Our results reproduce the edge-brightened structure up to $\sim1.3$ mas, i.e., in a more extended region compared with previous studies, with a brighter southern edge. The obtained synthetic image of the jet reproduces additional observed knots along the southern edge. 

\end{abstract}

\section{Introduction}
One of the best-studied objects across the entire electromagnetic spectrum is the radio galaxy M\,87, from radio wavelengths \citep{Kim2018a,Walker2018,Lu2023} up to $\gamma$-rays \citep{Abramowski2012,MAGIC2020}, and including optical \citep{Perlman2011} and X-ray bands \citep{Snios2019}. M\,87 hosts an edge-brightened jet, resolved with Very Long Baseline Interferometry (VLBI) observations. Nevertheless, its origin and formation mechanisms remain uncertain. Event Horizon Telescope (EHT) observations of its centre, designated M\,87${^\star}$, have revealed a ring-like structure that agrees with theoretical models of accretion onto a rotating Kerr black hole (BH) \citep{EHT_M87_PaperV,EHT_M87_2018}. The BH spin is considered a plausible energy source for the launching mechanism~\citep{Blandford1977,Moderski1996}, while magnetic fields are thought to be a key factor in the formation of relativistic jets \citep[e.g.][and references therein]{Tchekhovskoy2010}. M\,87 is located at a distance of $16.8\pm 0.8$\,Mpc \citep{Cantiello2018}, and the supermassive BH (SMBH) at its centre has an estimated mass of $(6.5\pm0.7)\times10^9\ M_\odot$ \citep[e.g.,][]{EHT_M87_PaperVI}, in agreement with previous estimations \citep[see, e.g.][]{Gebhardt11,Osorno2023}. 

Several studies of the morphological properties of the M\,87 jet have been conducted since its discovery. In these studies, different wavelengths and spatial scales were considered, particularly in the radio bands, where more scales can be observed \citep[e.g.][]{Algaba2021, Nikonov2023,Park2026}. The jet extends over 60\,kpc in length and shows a system of multiple knot-like features. An active feature is present at a projected distance of $\sim70$\,pc from the core, so-called HST-1 \citep{Birreta1999,Asada2012}. In addition to the large-scale morphology, recent and long-term observations across different radio bands have provided details about the jet structure near the central engine, where an edge-brightened jet has been detected \citep[e.g.][]{Walker2018,Lu2023,Kim2025}.

To properly test the models of the emission processes from M\,87 and study its Spectral Energy Distribution (SED), it is necessary to have an extensive, quasi-simultaneous multi-wavelength monitoring campaign of the source. However, these kinds of campaigns are complex because multiple facilities are involved, and in many cases, scheduling challenges arise. Nevertheless, in this context, two multi-wavelength campaigns in 2017 and 2018 were organised to study M\,87. These campaigns were an effort made by several groups to provide quasi-simultaneous observations for the Event Horizon Telescope (EHT) collaboration. In 2017, \citet{Algaba2021} captured M\,87 in a historically low state, and later in 2018, \citet{Algaba2024} detected the first VHE $\gamma$-ray flare from M\,87 since 2010.

Since we are also interested in studying the M\,87 jet at scales of milli-arcseconds (mas), VLBI observations are necessary. The M87 jet has been observed by several authors \citep[e.g.,][]{Kim2018a,Lu2023,Kim2025}. For this work we use the morphology of M87 jet at 86 GHz reported in \citet{Kim2018a}. They present data of M\,87's jet at 86 GHz (3.5\,mm) from 2004 to 2015 taken with the Global mm-VLBI Array (GMVA). From the set of images presented in their work, we chose one from 2014 February 26 to compare with the synthetic images produced by the general relativistic radiative transfer (GRRT) calculations, as it shows a smoother intensity map. 

Different theoretical models and simulations have been used to fit the emission properties and the morphology of AGN jets. The most common model used to explain the emission properties of jetted AGN is the so-called one-zone model \citep[e.g.,][]{Ghisellini1998,Celotti2008,Luna2024}. In this semi-analytical approach, a parameter exploration using Markov-chain Monte Carlo methods is performed to fit the multi-wavelength SED \citep[e.g.,][]{Tavecchio2008}.
This approach has been very successful in modelling the multi-wavelength SED of jetted AGN at different activity levels, including the emission produced in regions from pc to kpc-scales \citep[e.g.,][]{Ghisellini2010,Botcher2013}.

A different approach to model the jet emission, although restricted to scales very close to the SMBHs is done through the use of general relativistic magnetohydrodynamics (GRMHD) simulations \citep{DeVilliers03b,Gammie03} coupled with GRRT simulations \citep{Younsi2012,Pandya2016,Davelaar2019}. In this case, we start with a rotating black hole surrounded by a magnetised plasma torus. With a perturbation, the material in the torus begins to accrete towards the black hole and forms an accretion disc and a relativistic jet aligned with the spin of the black hole. Once the plasma has evolved, a radiative transfer is applied to the system to obtain the emission. In radiative transfer calculations, relationships between plasma properties and the emission and absorption coefficients are utilised \citep{Moscibrodzka2016}. Up to now, the GRMHD-GRRT models have reproduced the SED in the low-frequency regime ($10^{10}\ \text{Hz}\leq\nu\leq 10^{16}\ \text{Hz}$) but fail to reproduce the observed morphology, especially some of the characteristic features on the sheath and the edge-brightened jet \citep[e.g.][hereafter, Paper I]{Osorio2022} \citep[and][]{Fromm2021b, Zhang_M_2024,Yang2024}. {Although alternative approaches have been proposed \citep[e.g.,][]{Tsunetoe2025,Tsunetoe2026}, several issues remain unresolved that cannot be adequately addressed within these alternative scenarios.}

In this work, we introduce a new GRRT scenario considering a blob within the jet structure to improve the fit of these simulations to the observations of the M\,87 jet at sub-pc scales near the BH. In the Section \ref{sec:simul}, we describe the GRMHD and GRRT simulations used as a base for this work, while in Section \ref{sec:model} we present our new model and a discussion of the results. Finally, in Section \ref{sec:conc} we present the summary and conclusions.

\section{General relativistic radiative transfer}\label{sec:simul}

The multi-band emission calculations were performed in two steps. 
The first stage consists of GRMHD simulations of a magnetised 
accretion disc around a Kerr black hole spacetime, carried out 
using the code \texttt{BHAC} \citep{Porth2017,Olivares2019}. 
The initial hydrodynamic equilibrium torus is constructed 
following the Fishbone--Moncrief prescription, with a single-loop 
magnetic field introduced ad hoc, assuming the gas pressure 
to be 100 times larger than the magnetic pressure, corresponding 
to a plasma beta of $\beta = 100$. We consider a constant specific 
angular momentum values of $l\mathbf{:=-u_\phi/u_t} = 6.92,\, 6.84,$ and $6.76$. 
The size of the torus is defined by the inner edge at 
$r_{\rm in} = 20\,M$ and the pressure maximum at $r_{\rm c} = 40\,M$, where $M$ is the mass of the central BH \citep{Fishbone76,Font02b,Shiokawa2012,Rezzolla_book:2013,Cruz2020}. 
The plasma is modelled as a relativistic ideal gas with an 
adiabatic index $\Gamma = 4/3$ \citep[e.g.,][]{Rezzolla_book:2013}. The spherical-polar spatial coordinates $(r,\theta,\phi)$ span the simulation domain over the ranges $r\in[0.8r_\text{EH},2500\ M]$, $\theta\in[0,\pi]$, and $\phi\in[0,2\pi]$, where $r_\text{EH}=M+\sqrt{M-a_\star^2}$ denotes the event-horizon radius, with $a_\star$ the BH spin. The innermost cell wall lies inside this horizon. The radial grid spacing is logarithmic, while the two angular directions use uniform spacing. Three refinement levels are employed, yielding an effective grid resolution of $(N_r,N_\theta,N_\phi)=(384,192,192)$.

The second step consists of GRRT calculations to compute the 
multiband emission and radio images of the relativistic jet. 
The radiative transfer equations are solved using the 
\texttt{BHOSS} code \citep{Younsi2012,Younsi2020}. We perform the GRRT calculations using 200 snapshots of the evolved magnetised tori and relativistic jets obtained from the GRMHD simulations (see Section \ref{sec:model} and Table
\ref{tab:explored_parameters} for details of the adopted parameters). In the radiative transfer calculations, we restrict our treatment to synchrotron radiation, as it constitutes the dominant emission mechanism responsible for the so-called first hump of the SED of jetted AGN. Furthermore, we assume a non-thermal electron distribution function, since the radiation from plasma in the vicinity of a black hole cannot be fully explained by a purely thermal distribution \citep[Paper I;][]{Davelaar2018,Cruz2026,
Fromm2021b,Zhang_M_2024}. The non-thermal distribution consists of a thermal core at low energies combined with a power-law tail at higher energies, commonly referred to as a $\kappa$ distribution function \citep{Xiao2006,Davelaar2018,Ball2018}. 
The electron temperature is computed using the R-$\beta$ prescription \citep{Moscibrodzka2009}.
For the GRRT calculations, we use a FoV of $10^3$ M ($\sim4$ mas), the M87${^\star}$ black hole mass of  $6.5\times10^9\ M_\odot$ and a distance to the source of $16.8$ Mpc.

\section{A relativistic blob in a one-zone model}
\label{sec:model}

In many observations of relativistic jets in AGN, brightness enhancements moving along the jet are detected across different wavelengths and spatial scales \citep{Park2024,Bogensberger2024,An2013}. 
These features are commonly referred to as knots. In the case of M\,87, a prominent knot-- HST-1 --has been observed at a projected distance of $\sim 70$ pc from the core \citep{Harris2006}. 
Although HST-1 is not the only knot in the jet of this radio galaxy, it is the most extensively studied due to its proximity to the nucleus. {There are different works where this component has been considered a stationary reconfinement shock structure \citep[see for example][]{Stawarz2006,Mizuno2015}, however, there are other works that found that it could be an effect of the resolution in observations. \citet{Giroletti2012} found in 26 VLBI observations, obtained between 2006 and 2011, that HST-1 is resolved in complex substructures, with two main components that are found to move with apparently superluminal velocity ($\sim 4c$). In this study, we adopt the latter scenario and employ HST-1 as a representative case of the jet knots that motivate the present investigation.}

Many studies have been done to fit the multiwavelength SED of jetted 
AGN adopting the so-called one-zone model. In this approach, a single, spherical region within the jet is assumed to be the primary source of radiation. 
This emission, together with contributions from the accretion disc and other components of the AGN environment, produces the observed spectrum \citep[e.g., see][and references therein]{Luna2024}.

On the other hand, recent advances in kinetic plasma theory indicate that turbulent plasmas can give rise to long-lived plasmoids \cite{Imbrogno2024,Imbrogno2025}. In agreement with these findings, both ideal and resistive 
GRMHD simulations of magnetised accretion onto black holes show the self-consistent emergence of plasmoids within the accretion flow and 
the jet funnel. These simulations reveal the development of current sheets and a highly turbulent plasma morphology, capturing key features 
of magnetic reconnection and jet dynamics. In this context, plasmoids arise naturally as a consequence of turbulence and magneto-rotational instabilities. Those that are advected within the jet may constitute a plausible physical origin for the observed knots at larger distances from the black hole \citep{Chatterjee2019,Nathanail2020,Ripperda2020,Ripperda2022, Vos2024b,Jiang2024}. These plasmoids correspond to localized regions of enhanced temperature, which in turn give rise to locally increased emissivity and absorptivity \citep{Meringolo2023}.
Given the observational evidence and support from numerical simulations, it is natural to consider scenarios in which plasmoids or blobs play a 
central role in the emission processes of relativistic jets.

In this work, we introduce an ad hoc single blob-like structure within GRRT simulations to assess the impact of such structures on the broadband spectrum and image morphology at sub-parsec scales in 
the vicinity of the supermassive black hole M\,87${^\star}$. We model the blob as a spherical overdensity initially located at a distance $r_0$ from the 
black hole and propagating along the jet axis with a constant velocity $v_0$. As it travels outwards, the blob expands, with its radius defined 
by the iso-surface at which the plasma magnetisation reaches a prescribed value, $\sigma_{\rm cut}$. The magnetisation is defined as $\sigma = b^2/\rho$, where $b^2$ is the magnetic field strength in the fluid frame and $\rho$ is the rest-mass density. This threshold is treated as a free parameter in the model.
We incorporate the contribution of the blob to both emission and absorption by prescribing Gaussian profiles for the corresponding coefficients in the 
GRRT simulations, given by
\begin{align}
    \label{ec:emis_hs}
    j_{\nu,\text{b}} &= j_\nu A\exp\left[{(\vec{r}-\vec{r}_z)^2}/{2r_\text{b}^2(\sigma)}\right], \\
    \label{ec:abs_hs}
    \alpha_{\nu,\text{b}} &= \alpha_\nu A\exp\left[{(\vec{r}-\vec{r}_z)^2}/{2r_\text{b}^2(\sigma)}\right],  
\end{align}
where the blob center is located at $\vec{r}_\text{z}$ and $r_\text{b}$ 
denotes its radius, which depends on the magnetisation parameter 
$\sigma=\sigma_{\rm cut}$; 
$\vec{r}$ and $\vec{r}_z$ are three-vectors, and $A$ is an amplitude factor. 

In this framework, the blob contribution can be interpreted as a localised enhancement of both emission and absorption within a region of the jet. Assuming a constant velocity along its trajectory, the blob's motion is scaled using the Keplerian angular velocity, defined as
\begin{equation}
	\label{ec:Kep_vel}
    \Omega_K^\pm(r)=\pm\frac{{M^{1/2}}}{{r^{3/2}\pm a_{\star}M^{1/2}}},
\end{equation}
where $a_{\star}$ is the dimensionless BH spin, $M$ is its mass and $r$ is the radial distance to the black hole~\citep{Rezzolla_book:2013}.

To estimate the constant velocity, we define its magnitude as the product of the Keplerian angular velocity and the initial radial distance of the blob. We impose an upper limit on this velocity based on the maximum bulk velocity of the plasma within the jet spine, capping it at approximately eight times the initial Keplerian value. Because $\Omega_K^\pm(r)$ decreases monotonically with distance, a blob originating closer to the black hole naturally attain higher velocity than one formed further out. Given the velocity $v_0(r_0,t_0)$ at initial position, $r_0$, we compute the position of the blob, $\vec{r}_z$, at any time $t$ as
\begin{equation}
    \label{ec:plasmoid_pos}
    \vec{r}_z(t) = (0,0,r_0[v_0(t-t_0)+1]),
\end{equation}
where $\Omega_K$ is multiplied by $r_0$ to convert the angular velocity into a tangential velocity, given that  $v_0$ is expressed as a multiple of $\Omega_K$.

\begin{figure}
    \centering
    \includegraphics[width=1\linewidth]{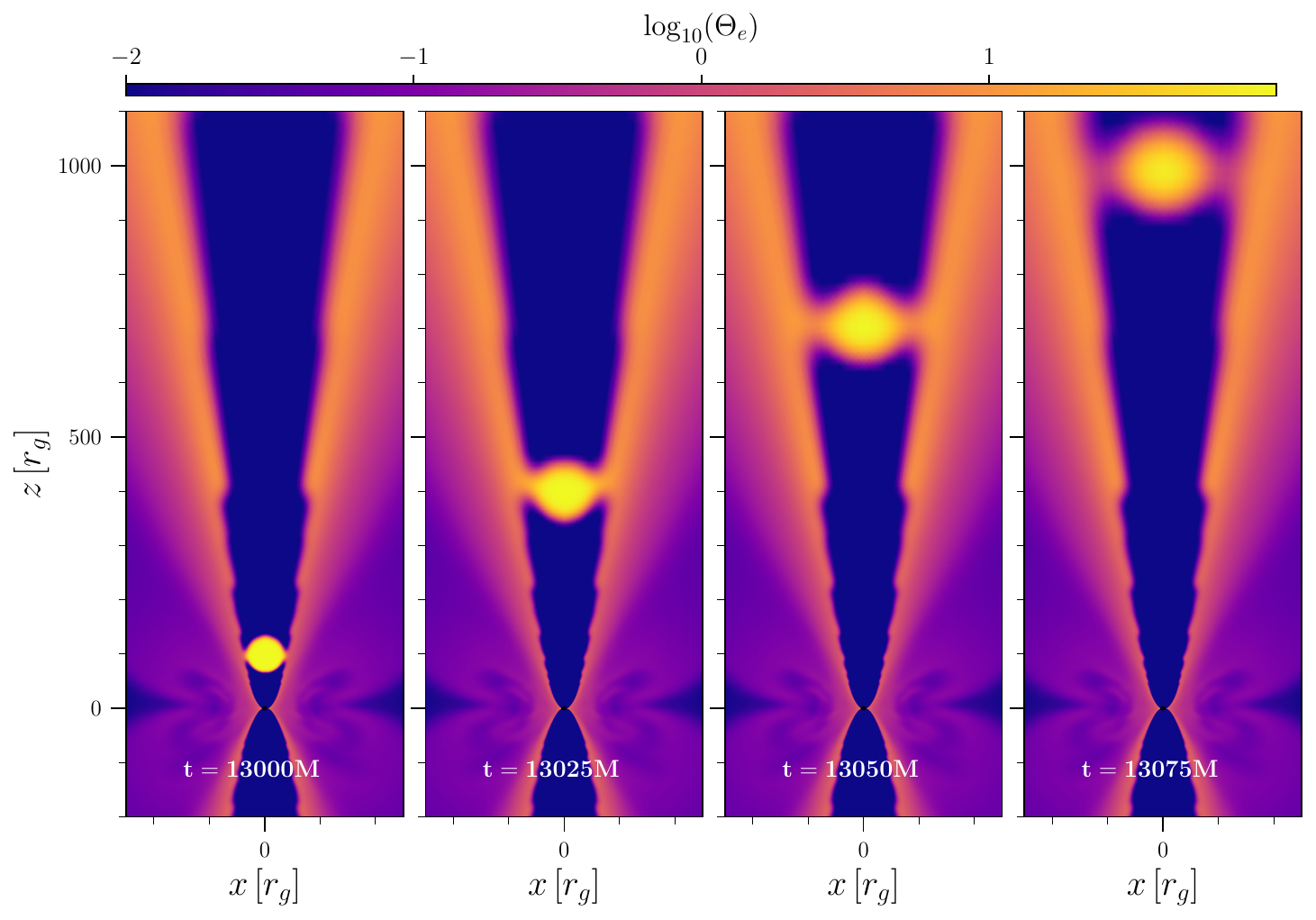}
    \caption{Dimensionless electron temperature of the plasma computed considering the R-$\beta$ prescription at different times. Initially, a spherical region with a higher temperature in the jet spine was added using magnetisation to compute the radius. Then, the sphere moves along the jet and increments its radius. This is just a schematic view of how the model is considered.}
    \label{fig:Pl_temp}
\end{figure}

Taking the jet axis as the reference frame for determining the blob radius, we employ the time- and azimuthally averaged plasma magnetisation to construct iso-contours associated with a prescribed magnetisation threshold, $\sigma_{\rm cut}$. These contours provide a relation between the position along the jet, $\vec{r}_z$, and the transverse coordinate $x$ at which the magnetisation reaches the $\sigma_{\rm cut}$. We then define this $x$ value as the blob radius at position $\vec{r}_z$, see Figure \ref{fig:Pl_temp} for an example of the blob evolution. Our GRRT calculations involve two sets of parameters: those describing the spacetime and plasma properties, and those characterising the relativistic blob. The former includes the black hole spin, $a_\star$; the electron temperature parameters, $R_\text{high}$ and $R_\text{low}$; the fraction of magnetic energy transferred to the electrons as kinetic energy, $\varepsilon$; the jet spine boundary, which also acts as a cutoff in the GRRT calculations, $\sigma_\text{cut}$; and the injection radius beyond which electrons are energised by the magnetic field, $r_\text{inj}$ \citep[see][]{Fromm2021b}.

The blob parameters, on the other hand, are its radius, $r_\text{b}$, determined by the magnetisation; its velocity along the $z$-direction, $v_0$; the amplification factor of the background emission, $A$; and its initial distance from the black hole, $r_0$. Given the large number of parameters involved in GRMHD and GRRT simulations, a comprehensive exploration of the parameter space would be required. However, due to the high computational cost of each simulation, we restrict our study to a reduced subset of parameters following previous studies by 
\citet{Fromm2021b}. This approach still allows for an adequate and scientifically meaningful exploration while keeping the computational expense manageable. 
The parameter values explored in this work are listed in 
Table \ref{tab:explored_parameters}, although only the best fit set of parameters is showed. The parameters on the left have been 
extensively studied by \citet{Fromm2021b}; therefore, we focus on those on 
the right, which are associated with our relativistic blob in a 
one-zone model.

\begin{table}
    \centering
    \caption{Model parameters and their explored values along this work.}
    \label{tab:explored_parameters}
    \def\arraystretch{1.1}
    \begin{tabular}{cc|cc}
        \hline
        Spacetime and & Values & One zone & Values \\
        plasma parameter &     & blob &  \\ \hline\hline
        $a_\star$ & $15/16$ \\
        $R_\text{high}$ & $80,\ 160,\ 240$ & $\sigma(r_\text{b})$ & $0.01,\ 0.1,\ 1$ \\
        $R_\text{low}$ & $0.1,\ 1,\ 10$ & $v_0\ [\Omega_K]$ & $2,\ 4,\ 5$ \\
        $\varepsilon$ & $0.0,\ 0.5$ & $A$ & $10,\ 50,\ 100$ \\
        $\sigma_\text{cut}$ & $1,\ 3,\ 5$ & $r_0\ [M]$ & $10,\ 100,\ 200$ \\
        $r_\text{inj}\ [M]$ & $10,\ 100,\ 200$ \\\hline
    \end{tabular}
\end{table}

We first calibrate the accretion rate unit, $\dot{M}_\text{unit}$, by matching the observed flux at 230 GHz. Specifically, we adopt the  value of $S_{230\,\text{GHz}} = 0.8$ Jy, consistent with the mean flux reported by \citet{Algaba2021, Algaba2024}. The optimal value of 
$\dot{M}_\text{unit}$ is determined using a bisection method, iterating until the mean total flux of the simulated images matches the observed value. At each step, we perform GRRT calculations over 200 snapshots of the GRMHD simulation from $13{,}000\,M$ to $15{,}000\,M$, where $M$ is the black hole mass. Once the flux at 230 GHz is reproduced, we use the calibrated $\dot{M}_\text{unit}$ to compute the spectral energy distribution over the frequency range $10^{10}\,\mathrm{Hz} \leq \nu \leq 10^{16}\,\mathrm{Hz}$. 
is done by averaging the total flux at each frequency across all snapshots. The observed SED of M87 exhibits an observed flux cut-off at approximately $\nu = 10^{15}\,\mathrm{Hz}$. \citet{Davelaar2019} proposed three possible physical origins for this feature: (i) synchrotron cooling in the jet, (ii) the emergence of a spectral break at the synchrotron burn-off limit, and (iii) a maximum energy constraint set by the Hillas criterion \citep{Hillas1984}. Guided by their analysis, we incorporate a cut-off at this characteristic frequency into our simulations. Additionally, we generate synthetic images at 86 GHz by convolving the GRRT images with the telescope beam, enabling direct comparison with the observations of \citet{Kim2018a}.

We apply observational constraints to our models by computing the 
$\chi^2$ statistic using the observed SED fluxes. This procedure is 
performed independently for both multiwavelength datasets, corresponding to the 2017 \citep{Algaba2021} and 2018 \citep{Algaba2024} campaigns. The resulting $\chi^2$ values are reported in columns (12) and (13) of Table \ref{tab:statistics_fit_width} for the 2017 and 2018 observations, 
respectively. Since our goal is to identify a model that simultaneously reproduces both epochs, we adopt the average of these two $\chi^2$ values as a constraint criterion. This averaged quantity is listed in column (14) of Table \ref{tab:statistics_fit_width}.

We also analyse the jet diameter in the convolved images and compare 
it with the width inferred from observations. The jet diameter in the 
synthetic images is computed following the procedure described by 
\citet{Fromm2021b} and Paper I. For this comparison, we use a moving window of $\pm 50$ snapshots; the snapshot that best reproduces the observed diameter and SED is reported in Table \ref{tab:statistics_fit_width}. The corresponding reduced $\chi^2_{50}$ values for the jet diameter of each model are listed in column (16). In addition, we morphologically identify the model that best matches the 86 GHz radio observations presented by \citet{Kim2018a}. Based on all these criteria, the best-fitting model is \texttt{M87.jet.1}.

\begin{table*}
    \centering
    \caption{Plasma and relativistic blob parameters. All models assume a 
    black hole spin of $a_\star = +15/16$. Plasma parameters are listed in 
    columns (2–7), while blob parameters are given in columns (8–10). 
    Column (11) reports the accretion rate. Columns (12–14) show the 
    $\chi^2$ values of the SED for the 2017 and 2018 observations, along 
    with their average. Column (15) indicates the snapshot time that best 
    reproduces the jet diameter, and column (16) lists the corresponding 
    reduced $\chi^2_{50}$ computed over 50 snapshots. The first row corresponds 
    to the results of Paper I.}
    \label{tab:statistics_fit_width}
    \resizebox{\textwidth}{!}{
    \begin{tabular}{c|ccccc|cccc|c|ccc|cc}
    \hline
        \rule{0pt}{1.1em} Model & $R_\text{high}$ & $R_\text{low}$ & $\varepsilon$ & $\sigma_\text{cut}$ & $r_\text{inj}\ [M]$ & $\sigma(r_\text{Bl})$ & $v_0\ [\Omega_K]$ & $A$ & $r_0\ [M]$ & $\dot{M}$ [$M_\odot\ \mathrm{yr^{-1}}$] & $\chi^2_{2017}$ & $\chi^2_{2018}$ & $\chi^2_\text{avg}$ & $t\ [M]$ & $\chi^2_\text{50}$ \\ 
        (1) & (2) & (3) & (4) & (5) & (6) & (7) & (8) & (9) & (10) & (11) & (12) & (13) & (14) & (15) & (16) \\ \hline\hline
        \texttt{Cruz-Osorio} & $160$ & $1.0$ & $0.5$ & $3$ & $10$ & -- & -- & -- & -- & $1.06\times10^{-4}$ & -- & -- & -- & -- & -- \\
        \texttt{M87.jet.1} & $80$ & $10.0$ & $0.0$ & $5$ & $200$ & 1 & 4 & 10 & 100 & $2.97\times10^{-4}$ & $1.637$ & $3.109$ & $2.373$ & $13380$ & $1.661$ \\
        \texttt{M87.jet.2} & $80$ & $10.0$ & $0.0$ & $5$ & $200$ & $0.01$ & $4$ & $10$ & $100$ & $1.98\times10^{-4}$ & $3.446$ & $4.757$ & $4.101$ & $13310$ & $2.098$ \\
        \texttt{M87.jet.3} & $80$ & $10.0$ & $0.0$ & $5$ & $200$ & $0.1$ & $4$ & $10$ & $100$ & $2.42\times10^{-4}$ & $2.402$ & $3.829$ & $3.115$ & $14500$ & $1.292$ \\
        \texttt{M87.jet.4} & $80$ & $10.0$ & $0.0$ & $5$ & $200$ & $1$ & $2$ & $10$ & $100$ & $2.78\times10^{-4}$ & $1.925$ & $3.476$ & $2.701$ & $13340$ & $1.523$ \\
        \texttt{M87.jet.5} & $80$ & $10.0$ & $0.0$ & $5$ & $200$ & $1$ & $5$ & $10$ & $100$ & $2.73\times10^{-4}$ & $1.882$ & $3.324$ & $2.603$ & $14110$ & $1.459$ \\
        \texttt{M87.jet.6} & $80$ & $10.0$ & $0.0$ & $5$ & $200$ & $1$ & $4$ & $50$ & $100$ & $1.72\times10^{-4}$ & $1.578$ & $2.976$ & $2.277$ & $14550$ & $2.499$ \\
        \texttt{M87.jet.7} & $80$ & $10.0$ & $0.0$ & $5$ & $200$ & $1$ & $4$ & $100$ & $100$ & $1.24\times10^{-4}$ & $1.622$ & $2.996$ & $2.309$ & $14550$ & $2.760$ \\
        \texttt{M87.jet.8} & $80$ & $10.0$ & $0.0$ & $5$ & $200$ & $1$ & $4$ & $10$ & $10$ & $3.05\times10^{-4}$ & $3.183$ & $4.682$ & $3.933$ & $13760$ & $1.552$ \\
        \texttt{M87.jet.9} & $80$ & $10.0$ & $0.0$ & $5$ & $200$ & $1$ & $4$ & $10$ & $200$ & $3.11\times10^{-4}$ & $1.566$ & $3.011$ & $2.288$ & $13250$ & $1.688$ \\ \hline
    \end{tabular}
    }
\end{table*}

\begin{figure}
    \centering
    \includegraphics[width=\linewidth]{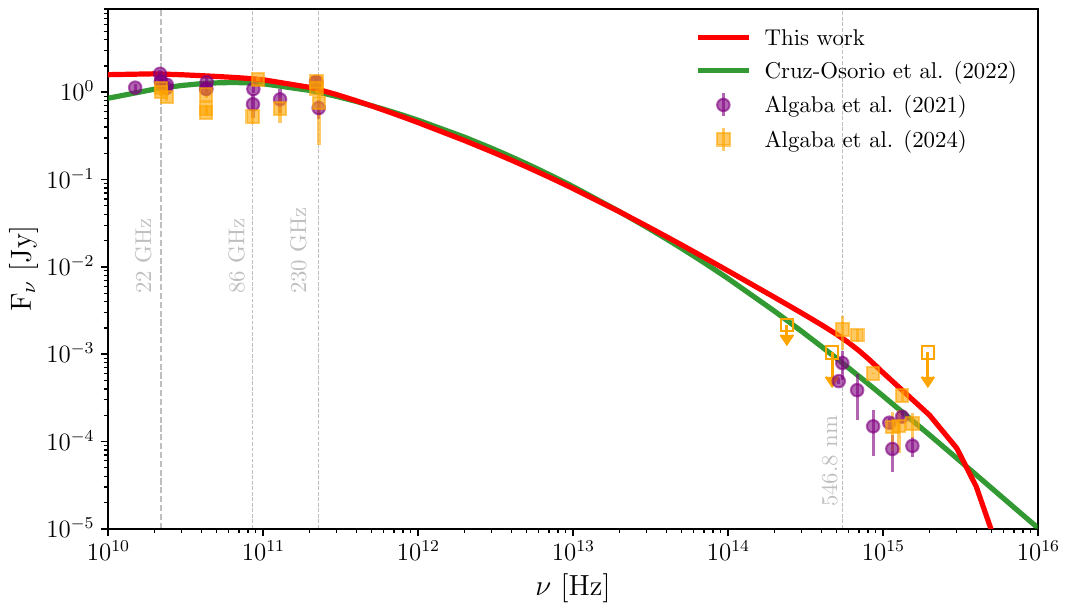}
    \caption{Spectral energy distribution. The red curve shows the results of this work, while the green curve corresponds to the model 
    from Paper I. Observational data from \citet{Algaba2021} and 
    \citet{Algaba2024} are shown in purple circles and yellow squares, 
    respectively.}
    \label{fig:Best_SED}
\end{figure}

\begin{figure*}
    \includegraphics[width=0.9\linewidth]{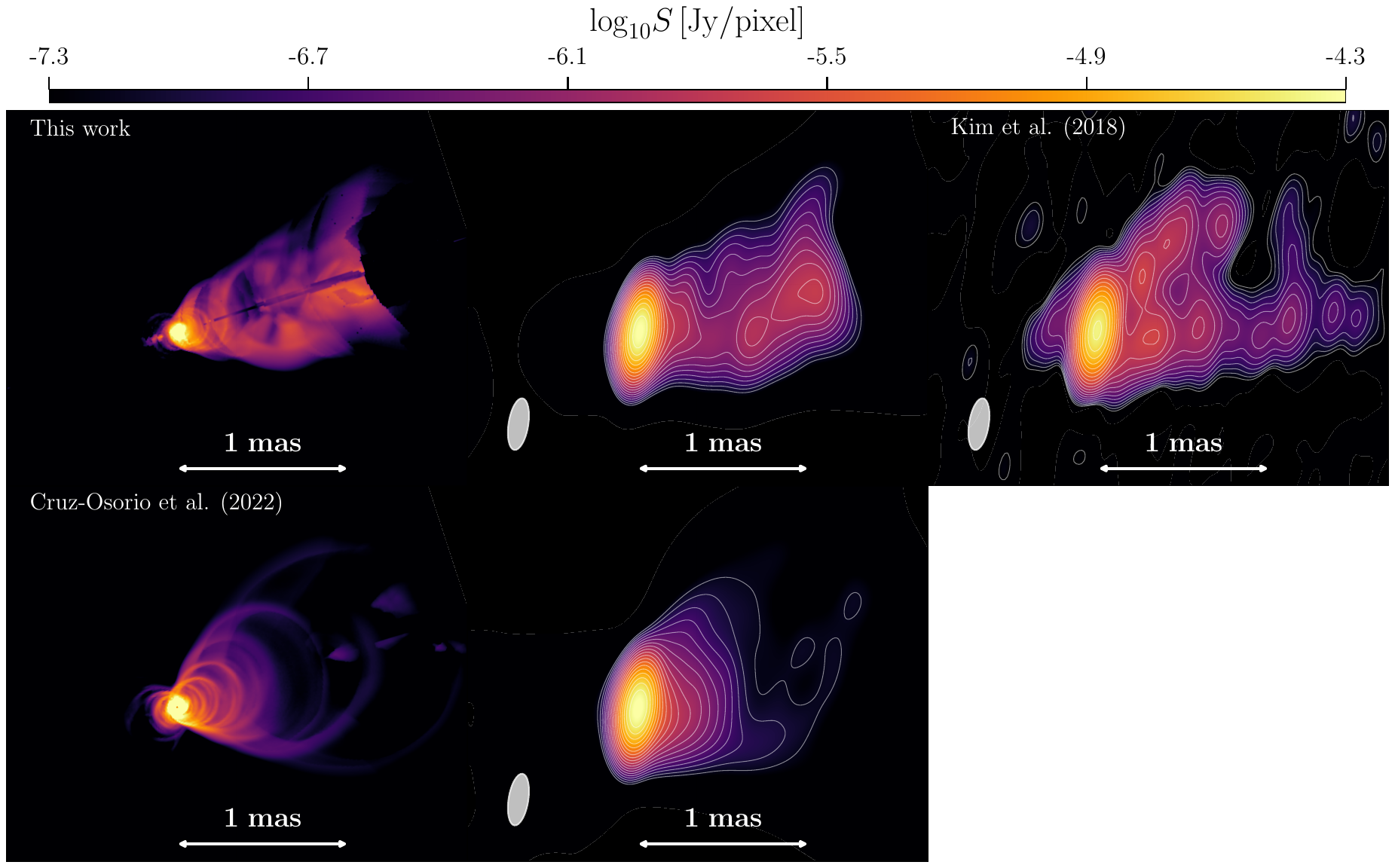}
    \caption{Morphology at 86 GHz for the best model presented in this work, compared with the results reported by Paper I. The left column displays the GRRT images at 86 GHz, the central column shows these images convolved with the GMVA beam, and the right column presents the observational data published by \citet{Kim2018a}. In both the observational and convolved images, the GMVA beam is represented by an ellipse.}
    \label{fig:Best_morphology}
\end{figure*}

\begin{figure}
    \centering
    \includegraphics[width=\linewidth]{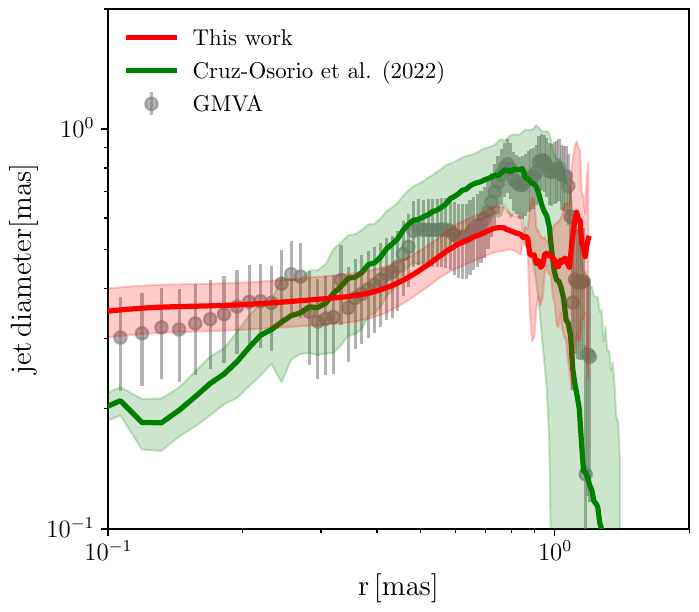}
    \caption{Jet diameter. The red curve 
    shows the results of this work, while the green curve corresponds to 
    the model from Paper I. Observational measurements from 
    \citet{Kim2018a} are shown as grey circles.}
    \label{fig:Best_width}
\end{figure}

In Figures \ref{fig:Best_SED}, \ref{fig:Best_morphology} and \ref{fig:Best_width}, 
we present the SED, morphology and jet diameter, respectively, for the best-fitting model, \texttt{M87.jet.1}. The results for this model are shown as red curves, while those corresponding to the best model from Paper I are displayed in green for comparison. In the SED panel, observational data from \citet{Algaba2021} and \citet{Algaba2024} are indicated by purple circles and yellow squares, respectively. In the jet diameter panel, the observational measurements from \citet{Kim2018a} are shown in grey.

In Table \ref{tab:statistics_fit_width}, we report the $\chi^{2}$ 
values of the SED for each model. Excluding the $\gamma$-ray flare, 
our model provides an improved fit to the optical/UV bands of the 2018 
observations. A key improvement in our SED modelling is the ability to 
reproduce a flatter spectrum in the radio band, which is characteristic 
of radio galaxies, particularly M\,87. Specifically, we obtain spectral 
indices of $\alpha_{22-86\,\text{GHz}} = 0.08$ in the radio band and 
$\alpha_{86-230\,\text{GHz}} = 0.29$ in the millimetre range. In contrast, 
previous studies have reported negative spectral indices in the radio, 
$\alpha_{22-86\,\text{GHz}} = -0.11$ and $\alpha_{22-86\,\text{GHz}} = -0.19$ \citep[Paper I,][]{Fromm2021b}.
Our results are closer to observational constraints, which indicate 
$\alpha_{22-86\,\text{GHz}} > 0.41$. These findings suggest that the 
inclusion of a blob component flattens the radio spectrum, while 
maintaining consistency with the EHT millimetre observations of the 
black hole shadow. 

In Figure \ref{fig:Best_morphology}, we present the 86 GHz morphology of the best-fitting model, compared with the results of Paper I and the GMVA observations reported by \citet{Kim2018a}. The convolved images are produced using the same GMVA beam, with a size of $116 \times 307\,\mu\mathrm{as}$ and a position angle of $-9^{\circ}$. Our model yields a more extended jet and reveals a distant transversal feature at around $1\,{\rm mas}$ consistent with a similar structure observed in the 86 GHz image. It also reproduces two bright, knot-like regions near the jet's southern edges, which are characteristic features of the M\,87 jet in GMVA observations. 
From the jet diameter measured in the best-fitting model, we find that the blob model provides a better fit to the data in the immediate vicinity of the black hole. This improvement can be attributed to the dominance of the blob component at those distances. Moreover, near the central engine, the higher magnetic field strength amplifies emission in the blob. Due to the image dynamic range, the blob emission outshines the extended emission so the jet width is dominated by the blob size. 
In the region between $0.2$ and $0.7$ mas, both models show good agreement with the observational data. However, reproducing the jet width across the full $0.1-1.0$ mas range remains challenging. Beyond $0.7$ mas, the blob model initially underestimates the jet width in the $0.8$–$1.0$ mas regime before recovering agreement at $\sim 1.2$ mas, corresponding to the transverse feature seen in the morphology. 
A deeper investigation of the jet structure will be necessary to fully understand the impact of these parameters in the blob model.

In summary, the best-fitting model consists of a rotating supermassive 
black hole with spin $a_\star = +15/16$ at the centre of the radio galaxy 
M\,87, that has a black hole mass of $6.5 \times 10^9,M_\odot$ \citep{EHT_M87_PaperVI} and located at a distance of $16.8$ Mpc from Earth \citep{Cantiello2018}. The plasma emission is modelled as nonthermal, with 
no direct contribution from magnetic energy ($\varepsilon = 0$), and electron temperature parameters $R_\text{high} = 80$ and $R_\text{low} = 10$, indicating slightly hotter electrons in the jet than in the disk. The jet spine is defined by the magnetisation contour $\sigma_\text{cut} = 5$, while magnetically energised 
particles are injected at $r_\text{inj} = 200,M$. The blob is characterised by 
a size determined at $\sigma = 1$, and propagates within the jet with 
velocity $v_0 = 4\Omega_K \sim 0.4c$, originating at $r_0 = 100,M$. It 
enhances the local emission and absorption by a factor $A = 10$. The resulting 
accretion rate is $\dot{M} = 2.97 \times 10^{-4}\,M_\odot\,\mathrm{yr}^{-1}$, 
consistent with previous observational estimates \citep{EHT_M87_PaperV}.

\section{Summary and conclusions}\label{sec:conc}

Ongoing and future observations of the M\,87 relativistic jet will 
significantly improve the characterisation of its morphology at 
sub-parsec scales, in the vicinity of the supermassive black hole. 
In this context, advanced theoretical and numerical modelling based 
on GRMHD and GRRT has become essential for understanding both jet 
launching mechanisms and the jet--disk connection. However, current 
state-of-the-art simulations of black hole--magnetised disk systems 
remain insufficient to fully reproduce the observed complexity of 
jet structures. To address this limitation, and to reproduce both 
the radio morphology at 86 GHz and the low-energy 
SED, we introduce a relativistic blob embedded within the jet spine 
in the context of a one-zone model. This blob propagates with 
super-Keplerian velocities and produces nonthermal emission, providing 
a physically motivated framework to capture key observational features.

We have carried out a systematic analysis of the impact of incorporating 
a relativistic blob, exploring a range of velocities, sizes, and 
emission amplification factors, and assessing their effects on both 
the jet morphology and the SED. As a result, our model \texttt{M87.jet.1} 
provides an improved SED, in particular by reproducing it with a flatter radio 
spectrum with a spectral index of $\alpha_{22-86\,\text{GHz}} = 0.08$, 
consistent with synchrotron self-absorption and characteristic of radio 
galaxies such as M\,87. Moreover, the model successfully captures key 
morphological features of the jet, including 
two knot-like regions along the southern edge, and a transverse feature 
at $\sim 1$ mas. These features are in good agreement with observations 
and arise naturally as a direct consequence of the presence of the blob 
within the jet. The measured jet width is sensitive to the dynamic range due to the fitting method employed. Within the available dynamic range, the localized brightness enhancement of the blob outshines the fainter background emission, resulting in a jet width dominated by the blob size in the region where it is located.

While our current model assumes equal amplification for emission and absorption, a blob with higher opacity would enhance synchrotron self-absorption. This produces a flatter radio spectrum and necessitates an 10\% increase in the accretion rate to match the observed flux level. From a morphological standpoint, however, the resulting jet structure would be very similar.

In summary, our results demonstrate that incorporating relativistic, 
nonthermal blobs is essential for better modelling the M\,87 jet. 
Their inclusion provides a natural explanation for both the observed 
spectral properties and the complex jet morphology. Further progress 
will require more comprehensive and self-consistent treatments, 
including the formation of blobs within GRMHD simulations, as well as 
the incorporation of polarimetric signatures and pitch-angle anisotropies. 
These developments will help in reproducing a more complete understanding 
of the jet-limb brightening observed in M\,87.

\section*{Acknowledgments}

This research was supported by Dirección General de Asuntos del Personal Académico-Universidad Nacional Autónoma de México grant IA103725 and by Ciencia B\'asica y de Frontera 2023-2024 program of SECIHTI M\'exico projects CBF2023-2024-1102. DM, EB and ACO acknowledge support from Dirección General de Asuntos del Personal Académico-Universidad Nacional Autónoma de México grant IN113026. Simulations were performed at Atocatl and Tochtli, LAMOD-UNAM cluster, and Miztli supercomputing cluster at Dirección General de Cómputo y Tecnologías de Información UNAM project LANCAD-UNAM-DGTIC-479. This research has made use of the NASA/IPAC Extragalactic Database (NED), which is operated by the Jet Propulsion Laboratory, California Institute of Technology, under contract with the National Aeronautics and Space Administration.

\section*{Data Availability}

The observations used are public, see \citet{Algaba2021}, \citet{Algaba2024}, and \citet{Kim2018a}. The data underlying this article will be shared on reasonable request to the corresponding author.





\bsp	
\label{lastpage}
\end{document}